\documentclass[runningheads]{llncs}
\usepackage{graphicx}
\usepackage{mathptmx}
\usepackage{xcolor}
\usepackage{tabularx} 
\usepackage{amsmath}  
\usepackage{graphicx} 
\usepackage{todonotes}
\usepackage{multirow}
\usepackage{paralist}
\usepackage{cleveref}[2012/02/15]
\usepackage{makecell}

\newcommand{\repeatthanks}{\textsuperscript{\thefootnote}}

\begin{document}
\title{$\bf \mu$-Net: A Deep Learning-Based Architecture for $\bf \mu$-CT Segmentation}

\author{Pierangela Bruno\inst{1}\thanks{First two authors contributed equally to this paper.} \and Edoardo De Rose\inst{1}\repeatthanks \and Carlo Adornetto\inst{1} \and Francesco~Calimeri\inst{1,5} \and Sandro Donato\inst{2,3,6}  \and  Raffaele Giuseppe Agostino\inst{2,6} \and Daniela Amelio\inst{4} \and Riccardo Barberi\inst{2,6} \and Maria Carmela Cerra\inst{4} \and Maria Caterina Crocco\inst{2,6} \and Mariacristina Filice\inst{4} \and Raffaele Filosa\inst{2,6} \and Gianluigi Greco\inst{1} \and Sandra Imbrogno\inst{4} \and Vincenzo Formoso\inst{2,6}}

\institute{University of Calabria, Mathematics and Computer Science, Rende, Italy 
\and University of Calabria, Department of Physics, Rende, Italy 
\and INFN, Frascati, Italy 
\and University of Calabria, DiBEST, Rende, Italy 
\and DLVSystem, Srl, Rende, Italy 
\and STAR Lab, Rende, Italy}

\maketitle              
\begin{abstract}
X-ray computed microtomography ($\mu$-CT) is a non-destructive technique that can generate high-resolution 3D images of the internal anatomy of medical and biological samples. These images enable clinicians to examine internal anatomy and gain insights into the disease or anatomical morphology. However, extracting relevant information from 3D images requires semantic segmentation of the regions of interest, which is usually done manually and results time-consuming and tedious. 
In this work, we propose a novel framework that uses a convolutional neural network (CNN) to automatically segment the full morphology of the heart of \textit{Carassius auratus}. The framework employs an optimized 2D CNN architecture that can infer a 3D segmentation of the sample, avoiding the high computational cost of a 3D CNN architecture. We tackle the challenges of handling large and high-resoluted image data (over a thousand pixels in each dimension) and a small training database (only three samples) by proposing a standard protocol for data normalization and processing. Moreover, we investigate how the noise, contrast, and spatial resolution of the sample and the training of the architecture are affected by the reconstruction technique, which depends on the number of input images. Experiments show that our framework significantly reduces the time required to segment new samples, allowing a faster microtomography analysis of the \textit{Carassius auratus} heart shape. Furthermore, our framework can work with any bio-image (biological and medical) from $\mu$-CT with high-resolution and small dataset size.

\end{abstract}

\section{Introduction}
X-ray Computed tomography (CT) is a powerful and widely used imaging tool that provides 3D digital gray-scale images of an object's internal structure; such images can be quantitatively analyzed to identify specific components of the 3D morphology. 
Modern CT is a valuable diagnostic tool that provides meaningful information reducing X-ray doses. 
$\mu$-CT is an even more powerful technique used in the study of human and animal anatomy in research and medicine~\cite{malimban2022deep}, allowing to achieve higher resolutions.
Computer-based approaches can ease and enhance the extraction of information and patterns from $\mu$-CT, leveraging, for instance, accurate semantic segmentation of the anatomical parts. 
Moreover, in the latest years, the use of image segmentation algorithms proved to be promising in facilitating analysis and detection of abnormalities~\cite{ross2023beyond,bruno2022assessing}. 
Those methods can be applied voxel-wise in a 3D context as well as pixel-wise, slice by slice~\cite{fu2021review}; however, 
instrumental noise, non-uniform intensity, and pixel discretization can limit the resolution of the image and 
obscure finer details. 
Traditional segmentation methods like thresholding and morphological filters are sensitive to parameter changes, leading to potential detail loss; conventional methods struggle with variations in phase/absorption contrast intensity~\cite{fu2021review}. 
Generally, 3D segmentation methods lack flexibility and adaptability, 
and determining the best method for a specific application is challenging, especially 
in medical imaging due to the heterogeneity of image characteristics and distributions~\cite{fu2021review}.
Deep Learning (DL) approaches such as CNNs 
rapidly became the state-of-the-art (SOTA) for medical image segmentation, classification, recognition, and report generation~\cite{litjens2017survey,ad10361320}, and have been widely applied in the field of CT.
However, few attempts have been made on $\mu$-CT; indeed, the wealth of information 
presents a significant challenge in terms of analysis and interpretation.
This is particularly evident in semantic segmentation tasks such as for 
kidney~\cite{DACRUZ2020103906}, cartilage~\cite{matula2022resolving}, temporal bone~\cite{nikan2020pwd}, lung~\cite{sforazzini2022deep} and thorax mouse $\mu$-CT~\cite{malimban2022deep}. 
Same applies to cardiac imaging, crucial 
for patient-specific intervention planning~\cite{clark2021advances}; here, primary datasets 
are mainly magnetic resonance imaging (MRI), but tomography datasets have started to be acquired, which have a higher resolution. 
Notable contributions include DL segmentation structure for MRI cardiac datasets~\cite{liu2020residual}, U-net variant for short-axis MRI~\cite{zheng20183}, DL approach for ECG-gated CT data~\cite{sharobeem2022validation}, novel pipeline for whole heart CT segmentation~\cite{xu2018cfun}. 
%

This work aims at defining a general framework for DL-based processing of high-resolution $\mu$-CT images in presence of small datasets, a prevalent scenario in the medical domain. 
The underlying rationale is that we can improve performance and reliability of image segmentation by considering each class individually. 
Our approach not only contributes at enhancing $\mu$-CT segmentation performances, but also fosters the usage of more lightweight models, in contrast to the current 
widespread usage of foundational models. 
The main contributions of this work can be summarized as follows. \\ 
    $-$ We build a new dataset consisting of $\mu$-CT images from \textit{C. auratus}'s heart, a teleost fish, also known as {\em goldfish}. \\
    $-$ We design a novel DL-based framework for extracting, enhancing, analyzing information from $\mu$-CT. 
    It extends SOTA semantic segmentation by defining  multiple models and ensembling strategies. 
    We present an implementation of the framework and assess it by designing and conducting an extensive experimental campaign over the newly introduced dataset.
    Results show that our proposal outperforms the SOTA  methods, exhibiting improved performance and reduced misclassification errors. \\
     $-$ We show how the application of 2D CNNs followed by custom post-processing to achieve 3D continuity reduces computational costs if compared with 3D CNNs. \\
    $-$ We design an approach for feasible and robust segmentation, explicitly suited for use cases in which a limited number of labeled samples is available. \\
    $-$ We study how the quality of 3D tomographic images affects architectural performance, given that obtaining such images requires to collect multiple projection images over a wide range of projection angles.
%
%

To the best of our knowledge, this is the first approach that proposes a combination of multiple DL-based models and a comprehensive ablation study to assess the benefits of different architectures and parameter components in the context of $\mu$-CT image segmentation, even with limited prior knowledge. 

\section{Proposed approach} \label{sec:prop_approach}
We present $\mu$-Net, a novel DL-based framework for the analysis and semantic segmentation of $\mu$-CT images. 
As already introduced, although the high resolution of $\mu$-CTs offers many advantages, images can either be too rich or have little variability between different tissues; this can negatively affect CNN generalization capability, resulting in misclassifications; this is further exacerbated when only a few images are available. 
The herein proposed framework addresses such challenges by automatically extracting meaningful information from $\mu$-CT images. 
$\mu$-Net faces different tasks with different specialized models; each model is trained to automatically solve a small part of the whole task: each model segments a different area of the heart, and we defined an ad-hoc ensembling procedure to combine the results. 
One of the key advantages of our approach is  versatility; indeed, our framework can be applied to any $\mu$-CT images in the medical domain, regardless of the organs/tissues involved,
thus
resulting as a valuable tool for researchers across various disciplines.
It is flexible and adaptable, as it can be configured with different architectures and customized according to the dataset; moreover, it is specifically tailored to preprocess and postprocess images of this kind with suitable filters, taking into account the 3D nature of the image. 
\begin{figure}[t]
\centering
\includegraphics[width=0.82\textwidth]{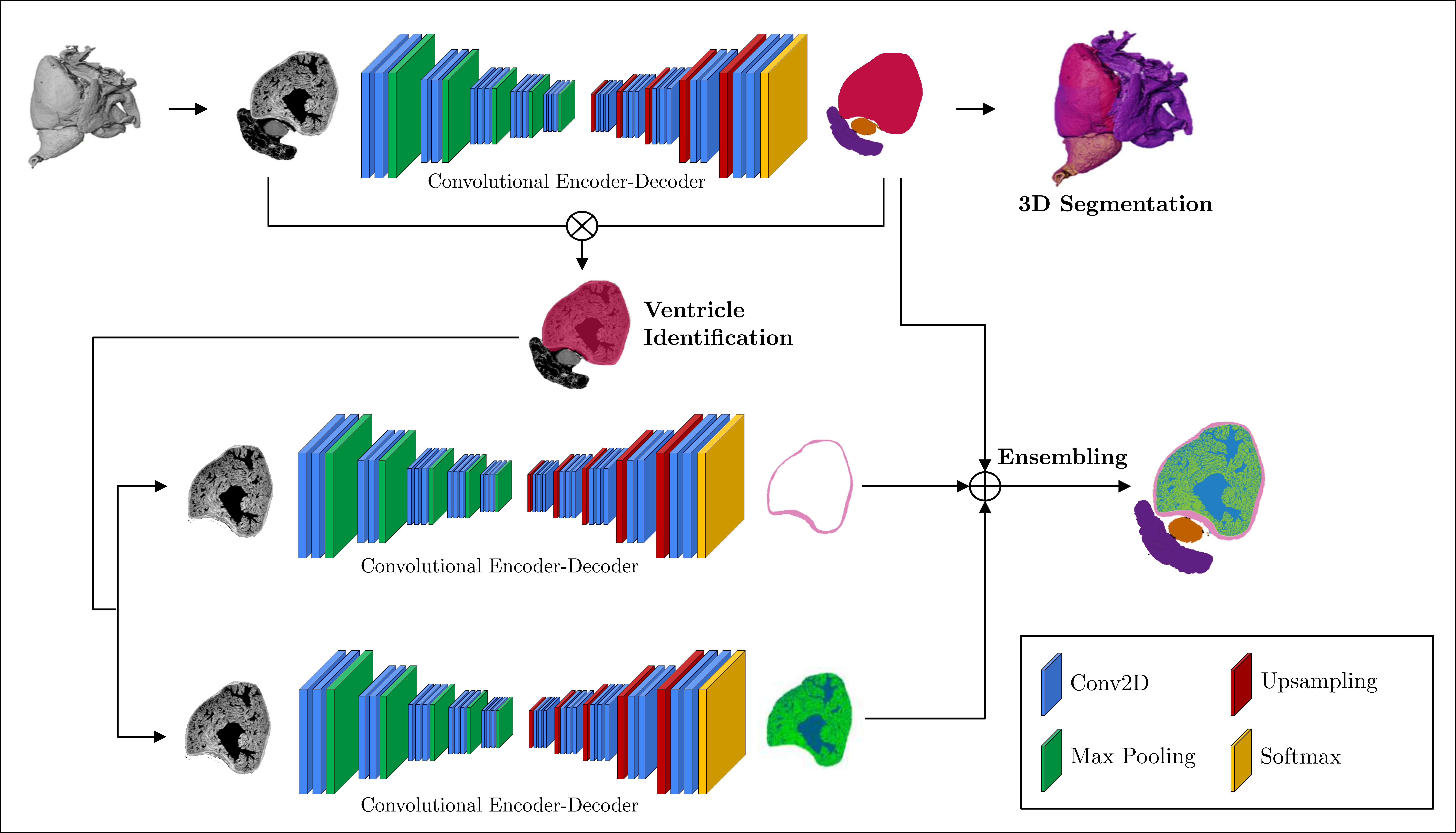}
\caption{Initially, the $\mu$-Net employs a CNN to identify the ventricles. Following this, two separate DL architectures carry out binary segmentation of various areas. The final result is obtained by applying an ensemble strategy to the different segmentations produced by each model.}
\label{fig:workflow_seg}
\end{figure}
Fig.~\ref{fig:workflow_seg} shows the architecture of $\mu$-Net whose aim is to automatically segment heart morphology in $\mu$-CT images of goldfish. We propose the following five-step procedure for performing semantic segmentation of the goldfish $\mu$-CT images: \textbf{\textit{1. Data acquisition and preparation}}: biological samples are collected and properly stained; subsequently, $\mu$-CT projections are acquired and the volume is reconstructed and normalized (see Section~\ref{sec:norm-descr}).\ \textbf{\textit{2. Data preprocessing}}: filters are applied to images according to the different tasks.\ \textbf{\textit{3. Segmentation model}}: we defined and trained three different models by dividing the segmentation problems (see Section~\ref{sec:training_phase});\ \textbf{\textit{4. Data post-processing}}: different filters and volumetric techniques are used to obtain a 3D coherence starting from 2D CNN models.\ \textbf{\textit{5. Ensembling models}}: we defined an ensembling set of rules to merge the results and obtain the whole final semantic segmentation (see Section~\ref{sec:training_phase}).

We experimentally validate the proposed methodology for the particular image segmentation task and compared the results with the SOTA methods (see Section~\ref{sub:sota}).
\section{Dataset building and description}\label{sec:norm-descr}
In our experiments, we use $\mu$-CT of the heart of \textit{C. auratus} (Linnaeus, C. (1758)), a teleost fish, also known as {\em goldfish}. 
The scans were manually annotated under the supervision of $4$ expert biologists for supervised learning.

The goldfish heart comprises four main components: sinus venosus, atrium, ventricle, and \textit{bulbous arteriosus}~\cite{filice2022goldfish}. 
Notably, the atrium features a spacious cavity with a muscular rim and a network of thin elastin and collagen fibers. 
Meanwhile, the ventricle consists of two distinct layers: the outer \textit{compacta}, rich in blood vessels and muscle bundles oriented in various directions, and the inner \textit{spongiosa}, which lacks blood vessels but contains numerous fibers.
%
Sample preparation and dataset acquisition consist of different steps; it is worth noting that the entire procedure requires several hours. 
Given the anonymous nature of the submission, additional details on data sources will be disclosed in case of publication.  
The X-ray acquisition resulted in a challenging procedure, as the biological nature of the samples posed several technical issues.
A staining procedure was applied to enhance the contrast of the samples; the absorption contrast technique was used to reconstruct 3D images. 
Each sample rotates with an angular step $\Delta\theta$, and is penetrated by an X-ray beam at each step; the attenuated X-ray beam's intensity is measured, creating a sinogram. 
The 3D structure is converted into a stack of 2D sinograms, that are fed to a reconstruction algorithm. 
We used the Filtered Back-Projection (FBP) algorithm for speed and simplicity. 
The resulting 2D image stack from FBP defines the 3D twin of the $\mu$-CT sample.
$\mu$-CTs are manually segmented to define the ground truth labels, 
which hold clinical significance and include atrium, ventricle, \textit{Bulbus arteriosus}, \textit{compacta}, and \textit{lacunary spaces}. 

The analysis of such dataset is significantly interesting for studying cardiac pathologies. 
When subjected to oxygen deprivation~\cite{filice2024functional,antiox9060555,IMBROGNO201424}, the goldfish accelerates its cardiac functions and, despite its small size, exhibits electrical activities similar to those of large mammals, which makes it relevant for translational research~\cite{bazmi2020excitation}.
Moreover, the goldfish heart is influenced by many hormones and peptides acting as cardiac modulators in mammals under normal and stressful conditions~\cite{imbrogno2019exploring}. 
These features make the goldfish an attractive model for exploring the mechanisms that give high flexibility to the heart, especially when facing internal and external challenges.

\section{Experimental Design}\label{sec:experiments}

\subsection{Data acquisition, 3D reconstruction and data preparation}\label{subsub:DA3R}
According to the previous section (see Sec.~\ref{sec:norm-descr}) and the difficulties that data acquisition and preparation hold, we acquired only 3 samples. 
For each sample, a total of $N_p = 3600$ projections with an angular resolution of $0.1$ degrees were acquired. Samples were reconstructed in line with the section~\ref{sec:norm-descr} using for each sample different projection dose. 
For each of the three samples, we used $N_p$, $\frac{N_p}{2}$, and $\frac{N_p}{3}$ projections to reconstruct three datasets, $D_1$, $D_2$, and $D_3$, respectively.
This allowed us to test how well our architecture can handle tasks with varying levels of projection data. We also conducted an analysis on the impact of input dimensions on the model’s performance. This was achieved by generating new, cropped stacks from the original data, focusing on the region of interest. This approach enabled us to evaluate the adaptability of our model to tasks with varying input sizes. Each reconstructed sample consists of approximately 1500 slices of size $1300 \times 1300$ pixels, where each voxel corresponds to $5,55\mu m$.
We implemented a normalization process for the dataset to ensure uniformity without introducing any additional bias. In the field of tomography, a specific range of absorption values is selected during the reconstruction phase. This selection process results in each sample appearing self-normalized.
For our 16-bit images, any pixel in the reconstructed image that falls below this threshold is assigned a value of 0, while those above it are assigned a value of $2^{16}$. Various strategies exist in the literature for this process, but we chose to reconstruct our samples using the same range of absorption values. This range serves as a normalization factor in our methodology.
Furthermore, in our experiment, we also considered the 2D-trans-axial projection of $\mu$-CTs: Axial View (corresponds to the XY plane, which is perpendicular to the rotation axis Z), Sagittal View (XZ plane), and Coronal View (YZ plane).
Then, we split the entire dataset of 9 stacks of images (3 datasets $\times$ 3 stacks), composed of $D_1$,\ $D_2$\ and $D_3$ into $D_{1,train}$,\ $D_{1,test}$,\ $D_{2,train}$,\ $D_{2,test}$,\ $D_{3,train}$ and $D_{3,test}$. 
In this setting, the train subsets consist of 2 stacks (2 heart samples). 
The third sample is used as a single heart sample that is common to all the datasets, except for the number of projections ($N_p$).
To reduce redundancy, we select only one image out of every three from the input stacks for training, since the images are nearly identical when they are next to each other.
For each training iteration, we randomly select one tile from each 2D slice to reduce its size. The training set is split into $70\%$ for training and $30\%$ for validation to monitor progress and prevent overfitting.

\subsection{Training phase and evaluation metrics}\label{sec:training_phase}
The framework was developed using Pytorch (v1.13.0). A high-performance computing node with two Tesla V100-PCIE-16GB GPUs, Intel® Xeon® Gold 5118 CPU (2.30GHz), and 512GB of RAM was used for training.
%
Jaccard index, also known as the Intersection Over Union (IoU) coefficient, is used as the evaluation metric during the training (i.e., 1 means perfect prediction, 0 worst prediction)~\cite{jaccard-dice-9116807,maier-hein2024metrics}. 
To assess the overall performance on the test set, after reconstructing the entire volume, we employed an IoU weighted by the frequency of each class in the 3D image. 
\subsection{Ablation study}\label{sub:ablstudy}
As for the ablation study we conducted, we explored: 
\begin{compactitem}
\item 
    \textbf{(A.1) Hyperparameter space:} learning rate, tile size, model architecture, preprocessing, and postprocessing. For each aspect, we compare several options and report the results on the validation set. 
\item 
    \textbf{(A.2) DL-based models: } Segnet~\cite{badrinarayanan2017segnet}, DeepLabV3~\cite{yurtkulu2019semantic} and U-net~\cite{ronneberger2015u}.
\item 
    \textbf{(A.3) Input parameters: } normalization type, tile size, number of slices for each stack (i.e., number of images are fed into the model at once), preprocessing and postprocessing methods. 
    To reduce the computational cost of processing large images and to avoid losing small-scale details relevant to our samples, we applied random cropping of sub-images, called tiles.
    We also experimented with different filters on input images (i.e., histogram equalization, median, unsharpmask filter).
\item 
    \textbf{(A.4) Number of projections:} variation of projection dose and, consequently, the generated 3D images (see Sec.~\ref{sec:norm-descr}); number of projections affects the quality of the image in terms of noise, spatial resolution, and artifacts. We found it useful to examine how performance changed with spatial resolution. 
\end{compactitem}
\subsection{Experiments}\label{subsec:exp} \label{sub:sota} 
As mentioned, we split the semantic segmentation task into separated sub-problems, focused on specific classes. Therefore, we have conducted several experiments: 
\textbf{(1) Semantic Segmentation of Atrium, Ventricle, and \textit{Bulbus arteriosus}}\label{subsub:segAVB}.\ 
The chosen model is the Segnet~\cite{badrinarayanan2017segnet} (see subsection~\ref{sub:ablstudy}). 
The train ran for $150$ epochs with a learning rate of $0.0001$, Adam optimizer, and tiles dimension of $400\times400$ pixels. We trained our model on the XY view of the sample and inferred on all three views (XY, XZ, and YZ). To obtain the final prediction and the 3D continuity, first, we chose the pixel-based mode for the pixels of intersection between the views and then we applied a hole-filling algorithm.\ 
\textbf{(2) Binary Segmentation of \textit{lacunary spaces}}.\  
We used the Segnet model with the same parameters as the previous step, except for the tile size of $224\times224$ pixels. The model received as input only the ventricle image part obtained from the previous step.
To enhance the contrast between \textit{lacunary spaces} and tissue, we applied an unsharp mask filter to the input image.\ 
\textbf{(3) Binary Segmentation of \textit{compacta}}.\ 
Similarly, we used a Segnet model with the same parameters and tile size. Also in this case we used as input only the ventricle part of the image.\ 
\textbf{(4) Overall Semantic Segmentation via Ensambling}.\ 
Once the results of the three experiments have been obtained, an ensembling strategy was performed to obtain a single segmentation result. Such strategy consists of a set of rules:
\textit{(I)} the Atrium class is always chosen over all other classes;
\textit{(II)} the \textit{Bulbus arteriosus} class is preferred over the \textit{lacunary spaces} and \textit{compacta} ones;
\textit{(III)} the \textit{compacta} class is selected in preference to the Ventricle and \textit{lacunary spaces} classes;
\textit{(IV)} the \textit{lacunary spaces} class is favored over the Ventricle class.\\ \\
\textbf{Comparison approaches}
To evaluate our approach, we compared it with two SOTA anatomical image segmentation tools: Biomedisa~\cite{LoeselVandeKampJayme2020_1000125797} and nnU-net~\cite{isensee2021nnu}. Biomedisa is a platform designed for semi-automatic and automatic segmentation of large volumetric images, using smart interpolation of sparsely pre-segmented slices. It should be highlighted that we made use of the automatic image segmentation version of Biomedisa and it takes in input 3D images. nnU-net, on the other hand, is a DL-based method that self-configures for any new segmentation task, covering preprocessing, network architecture, training, and post-processing.

\section{Results and Discussion}\label{sec:discussionandresults}
Among the tested architectures (see Sec.~\ref{sub:ablstudy} (A.2)) we discarded DeepLabV3 due to its poor accuracy. Although U-net achieved good performance, Segnet obtained the best results on the test set. Therefore, we selected Segnet as the model for each experiment.
Also, according to Sec.~\ref{sub:ablstudy} (A.3), we tested two data normalization techniques: $a$ normalize the data based on the mean and standard deviation of each stack and $b$ normalize during the reconstruction process. The technique described in $b$ yielded better results in the ablation study so we adopted it.
As stated by Sec.~\ref{sub:ablstudy} (A.4), we trained different models varying on the number of projections used for the reconstruction stack. 
\\
Results of the first experiment (see Sec.~\ref{subsec:exp}) show that as the number of projection doses decreases, the spatial resolution deteriorates \cite{villarraga2020effect}. The models were then evaluated on various test sets with different numbers of projection doses. Results are reported in Tab.~\ref{tab:IOU3parti}. 
The models were tested by using the test sets for each of the 3 different datasets and performing 3-fold cross-validation over the 3 stacks in each dataset. 
\begin{table}[t]
\centering
\caption{{Performance achieved by $\mu$-Net on the test set according to different training sets with a different number of projection doses.} }
\label{tab:IOU3parti}
\begin{tabular}{l@{\hskip 0.07in}|@{\hskip 0.07in}lll}
\multicolumn{1}{c}{\multirow{2}{*}{Train set}} & \multicolumn{3}{c}{IOU (\%) on Test set} \\ \cline{2-4} 
\multicolumn{1}{c}{}                           & $D_{1,test}$   & $D_{2,test}$  & $D_{3,test}$  \\ \hline
$D_{1,train}$                                     & \textbf{87.9  $(\pm 3.7)$ }     & 77.6  $(\pm 4.5)$     & 44.6  $(\pm 7.8)$     \\
$D_{2,train}$                                     & 44.5  $(\pm 7.8)$      & 73.7 $(\pm 3.4)$   & 81.1 $(\pm 3.5)$      \\
$D_{3,train}$                                     & 30.3  $(\pm 8.5)$      & 81.7 $(\pm 3.6)$      & 86.8  $(\pm 3.6)$     \\
$D_{1,train} + D_{2,train}$                         & \textbf{87.9  $(\pm 3.7)$}      & \textbf{86.4  $(\pm 3.4)$}     & \textbf{88.6 $(\pm 3.7)$}   
\end{tabular}
\end{table}

\begin{table}[t]
\centering
\caption{{Comparing IOU scores for our proposal, nnU-net, and Biomedisa methods. Best results for each class are reported in bold.}}
\label{tab:IOUtot}
\begin{tabular}{l@{\hskip 0.07in}|@{\hskip 0.07in}cccccc}
\multirow{2}{*}{} & \multicolumn{6}{c}{IOU (\%) on Test set} \\
\cline{2-7}
& Ventricle & \makecell{\textit{Bulbus} \\ \textit{arteriosus}} & Atrium & \textit{Compacta} & \makecell{\textit{Lacunary}\\ \textit{spaces}} & Total\\ \hline
$\mu$-Net & \textbf{94.5} $(\pm 3.4)$ & 77.5 $(\pm 3.7)$ & \textbf{87.9} $(\pm 2.4)$ & \textbf{77.2} $(\pm 2.4)$ & \textbf{84.8} $(\pm 3.6)$ & \textbf{87.6} $(\pm 3.7)$ \\
nnU-net & 80.2 $(\pm 3.4)$ & \textbf{80.1} $(\pm 3.4)$ & 61.1 $(\pm 5.2)$ & 64.8 $(\pm 4.2)$ & 72.9 $(\pm 4.4)$ & 76.8 $(\pm 5.2)$ \\
biomedisa & 63.5 $(\pm 5.4)$ & 16.3 $(\pm 8.4)$ & 20.5 $(\pm 7.4)$ & 50.2 $(\pm 5.6)$ & 56.9 $(\pm 5.4)$ & 43.8 $(\pm 8.4)$
\end{tabular}
\end{table}
The table shows that models trained with spatial high-resolution images obtain good performances on a spatial high-resolution test set, while worse performances are reported for a spatial low-resolution test set (see the first row of Tab.~\ref{tab:IOU3parti}). Training on spatial medium and low-resolution images results in better performance in similar-resolution test sets than the ones obtained on spatial high-resolution  (see the second and third row of Tab.~\ref{tab:IOU3parti}). This evidences that as the number of projections (and hence the spatial resolution) decreases, the performance of the network deteriorates due to a lack of information.
However, training a model using images with different spatial resolutions results in a more stable performance, meaning that the network has more generalization capability across resolutions. \\
Tab.~\ref{tab:IOUtot} shows performance results in terms of IOUs for each experiment performed (see Sec.~\ref{subsec:exp}). They largely hinge on the outcome of the initial experiment, the better the ventricle is detected in the first experiment the better the \textit{compacta} and \textit{lacunary spaces} will be detected in the same area of interest.
Our workflow achieved an IOU value of $87.6\%$ where a very good identification of ventricle ($94.5\%$) allows a good detection of \textit{compacta} and \textit{lacunary spaces}, $77.2\%$ and $84.8\%$, respectively.
We compared our proposal with two SOTA semantic segmentation models, nnU-net and Biomedisa. We trained both models to segment atrium, ventricle, and \textit{Bulbus arteriosus} regions only; our workflow achieved a higher IOU than both models, with 76.8\% for nnU-net and 43.8\% for Biomedisa.
The lower performance of nnU-net may be attributed to its automatic selection of patch size, filters, and normalization. We used the 2D configuration of nnU-net, as the 3D one was not feasible due to the limited number of samples (less than 3) and the high computational demand. On the contrary, Biomedisa’s performance was poor due to its use of a 3D U-net that standardized each sample size, leading to a reduction in resolution and distortion of shapes. 
\begin{figure}[t]
    \centering
\includegraphics[width=0.95\textwidth]{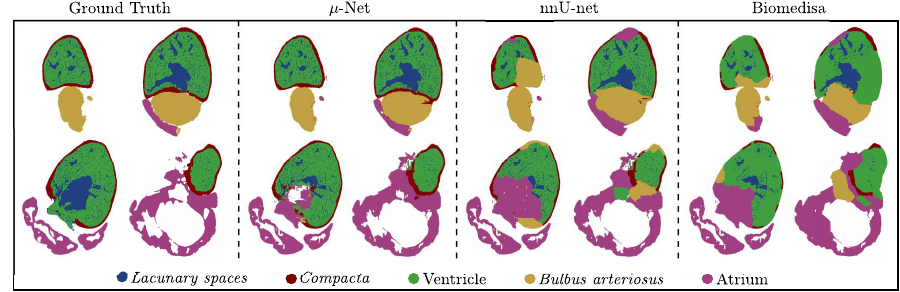} 
\caption{Visualization of ground truth, $\mu$-Net and the comparison methods results. Each image represents a slice taken from the same quartile of slices within a single 3D stack}
\label{fig:results}
\end{figure}
A visual inspection of the results is shown in Fig.~\ref{fig:results} where we compared from the left to the right: the manual segmentation (labels), the predicted segmentation of $\mu$-Net, and the results of the comparison methods. 
We can observe that $\mu$-Net demonstrates excellent segmentation of small and medium \textit{lacunary spaces}, while it tends to confuse larger ones with the background. As for other anatomical regions, $\mu$-Net notably outperforms the comparison methods.
In accordance with our hypothesis, these results highlight how the strategy underlying our framework, which decomposes the problem into simpler sub-problems before utilizing ensemble techniques, is more effective than tackling the problem as a whole, as seen in the cases of nnU-Net and Biomedisa.
Finally, our workflow can speed up the processing of new scans and is adaptable for additional segmentation tasks. The models are primed for training on more goldfish hearts or for transfer learning on other high-resolution $\mu$-CT tasks. The data from our automatic segmentation can be quantitatively analyzed like manually segmented data.

\section{Conclusions}

We introduced $\mu$-Net, an novel workflow built on Segnet for the semantic segmentation of  biological $\mu$-CT images. 
Training was performed on a new dataset, encompassing the collection of manually segmented 3D $\mu$-CT scans of goldfish hearts. 
The experiments showed that $\mu$-Net enhances efficiency and dependability of image segmentation techniques by treating each class separately. 
$\mu$-Net significantly outperformed existing methods, setting the stage for a more precise and automated examination of goldfish heart morphology and potential diseases, thus facilitating translational studies.

\bibliographystyle{splncs04}
\bibliography{bib}

\end{document}